\documentclass[11pt,a4paper,final]{article}
\usepackage[utf8]{inputenc}
\usepackage[english]{babel}
\usepackage{amsmath}
\usepackage{amsfonts}
\usepackage{amssymb}
\usepackage{appendix}
\usepackage{faktor}
\usepackage[hidelinks]{hyperref}
\hypersetup{
    colorlinks=true,
    urlcolor=blue,
    linkcolor=blue,
    citecolor=red,
    }
\usepackage{amsthm}
\usepackage{bbold}
\usepackage{mathrsfs}
\usepackage{wasysym}
\usepackage{mathtools}
\usepackage{bm}
\usepackage{cancel}
\usepackage{color}
\usepackage{enumitem,multicol}
\usepackage{multicol}
\usepackage{tikz}
\usepackage{tikz-cd}
\usepackage[Glenn]{fncychap}
\usepackage{subfig}
\usepackage{accents}
\usepackage{slashed}
\usepackage{psfrag}
\usepackage{stackengine}
	
\usetikzlibrary{babel}
\makeatletter
\newcommand*\owedge{\mathpalette\@owedge\relax}
\newcommand*\@owedge[1]{%
	\mathbin{%
		\ooalign{%
			$#1\m@th\bigcirc$\cr
			\hidewidth$#1\m@th\wedge$\hidewidth\cr
		}%
	}%
}
\makeatother

\usepackage{cjhebrew}

\newcounter{mnotecount}

\newcommand{\mnote}[1]
{\protect{\stepcounter{mnotecount}}$^{\mbox{\tiny
			$\,\bullet$\themnotecount}}$ \marginpar{
		\raggedright\tiny\em
		$\,\bullet$\themnotecount: #1} }

\newtheorem{teo}{Theorem}[section]
\newtheorem{cor}[teo]{Corollary}
\newtheorem{prop}[teo]{Proposition}
\newtheorem{lema}[teo]{Lemma}

\usepackage{makeidx}
\usepackage{graphicx}
\usepackage[left=2.60cm, right=2.60cm, top=2.45cm, bottom=2.80cm]{geometry}

\usepackage{scalerel}
\usepackage{stackengine,wasysym}

\makeatletter
\newcommand{\douwidehat}[2]{%
	\sbox0{$\m@th#1\widehat{\hphantom{#2}}$}%
	\sbox2{$\m@th#1x$}
	\sbox4{$\m@th#1#2$}
	\dimen0=\ht0
	\advance\dimen0 -.8\ht2
	\dimen2=\dp4
	\rlap{%
		\raisebox{\dimexpr\dimen0-\dimen2}{%
			\scalebox{1}[-1]{\box0}%
		}%
	}%
	{#2}%
}
\makeatother

\usepackage{color}
    \makeatletter
\renewcommand\part{%
	\if@openright
	\cleardoublepage
	\else
	\clearpage
	\fi
	\thispagestyle{empty}%
	\if@twocolumn
	\onecolumn
	\@tempswatrue
	\else
	\@tempswafalse
	\fi
	\null\vfil
	\secdef\@part\@spart}
\makeatother

\newcommand{\scri}{\mathscr{J}}

\newcommand{\wt}{\widetilde}
\newcommand{\wh}{\widehat}

\newcommand{\sch}{\operatorname{Sch}}

\newcommand{\scal}{\operatorname{Scal}}

\newcommand{\mf}{\mathfrak}
\newcommand{\real}{\mathbb R}
\newcommand{\tr}{\operatorname{tr}}

\renewcommand{\div}{\operatorname{div}}
\newcommand{\grad}{\text{grad}}

\newcommand{\st}{\stackrel}
\newcommand{\hess}{\operatorname{Hess }}

\newcommand{\X}{\mathfrak{X}}
\renewcommand{\d}{\coloneqq}

\newcommand{\ric}{\operatorname{Ric}}

\newcommand{\lie}{\mathcal{L}}
\newcommand{\mc}{\mathcal}
\newcommand{\sto}{\rightarrow}
\renewcommand{\to}{\longrightarrow}

\renewcommand{\mapsto}{\longmapsto}
\title{\vspace*{-1.35cm}\textbf{Conformal characterization of the Fefferman-Graham ambient metric}}
\author{Marc Mars\footnote{\href{mailto:marc@usal.es}{marc@usal.es}}\,\, and Gabriel Sánchez-Pérez\footnote{\href{mailto:gasape21@usal.es}{gasape21@usal.es}} \\
	Departamento de Física Fundamental, Universidad de Salamanca\\
	Plaza de la Merced s/n, 37008 Salamanca, Spain}
\date{\today}

\begin{document}
	\maketitle
	\begin{abstract}
In this paper, we study the asymptotic structure of the Fefferman-Graham ambient metric. We prove that every straight ambient metric admits a conformal completion with a well-defined null infinity, and that the asymptotic expansion of the metric at infinity can be related to that at the homothetic horizon. Furthermore, in even dimensions, we show that the Fefferman-Graham obstruction tensor naturally arises in the geometry at infinity. By identifying the fundamental properties that this particular conformal extension exhibits, and analysing their sufficiency, we arrive at the main result of the paper, namely the identification of a set of conformally covariant conditions that completely characterize the ambient metric from a conformal perspective. In particular, our result relaxes the requirement of the homothety one-form being exact.
	\end{abstract}
	\section{Introduction}
	
Homothetic vector fields play a fundamental role both in general relativity and in conformal geometry. Mathematically, they generate scale transformations, providing a natural framework to study conformal invariants and related structures \cite{fefferman1985conformal}. In mathematical relativity, homotheties often arise as self-similar limits of families of solutions to Einstein’s equations, such as in Christodoulou's proof of cosmic censorship in spherical symmetry \cite{christodoulou1999instability}. From the physical side, they are central to the analysis of critical phenomena in gravitational collapse, where self-similarity governs the threshold between dispersion and black hole formation \cite{choptuik1993universality,gundlach2007critical}. It is therefore natural to investigate self-similar solutions of Einstein’s equations, with particular attention to their asymptotic behaviour.\\

In their seminal work \cite{fefferman1985conformal} (later expanded into the monograph \cite{fefferman2012ambient}), Fefferman and Graham developed a powerful formalism to study conformal invariants from a differential geometry perspective. Their construction is based on the observation that the light-cone of a point in Minkowski spacetime (which is a homothetic horizon) encodes the full conformal structure of the sphere, and seeks to extend this picture to arbitrary conformal classes. Beyond its intrinsic mathematical appeal, this formalism has become central in several physically motivated contexts such as holography (see \cite{parisini2024ambient} and references therein). The basic idea is to associate to each conformal structure $(\mc S,[h])$ of signature $(p,q)$ another ambient manifold $(\mc M,g)$ of signature $(p+1,q+1)$ satisfying the following properties: (i) $(\mc M,g)$ admits a homothetic horizon with homothety $T$; (ii) the horizon encodes the conformal structure $(\mc S,[h])$, in the sense that any of its sections is conformal to $(\mc S,h)$; and (iii) the ambient metric is Ricci-flat to all orders at the horizon.\\

Starting from a natural ansatz for the ambient metric, Fefferman and Graham write down the derivatives of its Ricci tensor in terms of the derivatives of the metric coefficients. They prove that, in odd dimensions, the Einstein equations uniquely determine the coefficients of the ambient metric as a formal power series to infinite order. In even dimensions, however, this recursive determination is only possible up to certain order, and an additional symmetric, traceless tensor must be specified as free data to continue the expansion. In this setting, the so-called obstruction tensor emerges, whose vanishing or not characterizes whether the ambient metric can be Ricci-flat to all orders or not. In both odd and even dimensions, the one-form $\bm T$ associated to the homothety is exact (as a one-form) up to the order determined by the equations, namely to infinite order in the odd case, and up to the order of the obstruction in the even case. Moreover, in even dimensions, if the free data satisfies a particular divergence condition, $\bm T$ becomes exact to infinite order as well. Ambient metrics satisfying these conditions are known as \textit{straight}.\\

There exist natural generalizations of the ambient metric beyond the smooth class. In the even dimensional case, and when the obstruction tensor does not vanish, one can enforce the Ricci tensor to vanish to infinite order by introducing logarithmic terms into the expansion that compensate the presence of the obstruction tensor, but at the same time spoil the smoothness of the ambient metric at the horizon. In odd dimensions, one can naturally consider solutions with expansions involving half-integer powers which also have an indeterminacy at a certain order.\\

The existence of ambient metrics is a intensively studied topic in the literature. For analytic data $(h,\Psi)$, standard convergence methods can be applied to establish existence of the ambient metric when the obstruction tensor vanishes (see \cite{baouendi1976singular}) and also when log terms occur \cite{kichenassamy2004conjecture}. In the general smooth case, the results in \cite{anderson2005existence,anderson2005asymptotically,kaminski2023well} can be used to prove existence of odd dimensional ambient metrics beyond the analytic case. In a remarkable breakthrough, the authors of \cite{rodnianski} study a class of metrics, called \textit{self-similar}, defined as metrics admitting a homothetic horizon and solving \textit{exactly} the Einstein equations. These metrics are obtained after defining characteristic initial data and applying a suitable scaling limit. One of the fundamental results in \cite{rodnianski} is the existence and uniqueness of a self-similar, regular solution realizing prescribed data consisting of a Riemannian metric and a symmetric traceless $(0,2)$ tensor. The precise notion of regularity in this construction depends crucially on the spacetime dimension.\\

In this paper, we study the Fefferman-Graham ambient metric, with particular emphasis on its asymptotic structure. We prove that every straight ambient metric admits a conformal completion with a well-defined null infinity, and that in even dimensions the Fefferman-Graham obstruction tensor naturally arises in the geometry at infinity, in the sense that the conformal Einstein equations fail to hold at infinity at a certain order of derivatives whenever the Fefferman-Graham obstruction tensor does not vanish. We further study the conformal properties of the ambient metric and identify a necessary and sufficient set of conditions that completely characterize it from a conformal perspective. Finally, we show that the requirement of the homothety one-form $\bm T$ being exact can be relaxed, thereby extending the scope of the construction.\\

Given a straight ambient metric $g$ in Fefferman-Graham coordinates, we construct a coordinate transformation and a conformal metric $\wt g$ that extends $g$ to null infinity $\scri$. This extension admits a bifurcate Killing horizon with integrable Killing one-form, with one branch corresponding to $\scri$ and the other to the original homothetic horizon. Moreover, the transverse derivatives of the metric at infinity coincide with the Fefferman–Graham expansion at the horizon. In even dimensions, we further determine the Fefferman-Graham obstruction tensor directly at infinity.\\

Having understood the basic features of this particular conformal completion, namely (i) the presence of a Killing field with bifurcate horizon, and (ii) its integrability, it is natural to ask for a set of sufficient conditions that determine the ambient metric from a conformal viewpoint. As we shall see, there are three. The first (I) is the existence of a conformal Killing vector admitting a bifurcate horizon, which is the natural conformally-invariant replacement for condition (i). The second (II) is condition (ii), which is already conformally invariant. And finally, we need to ask for an extra condition (III) relating the conformal Killing and the conformal factor in a conformally-covariant way. Our main result (Theorem \ref{teorema}) establishes that these three conditions fully characterize the ambient metric.

\begin{teo}[Informal version]
Let $(\mc M,g,\Omega)$ be a regular conformal spacetime satisfying the conformal Einstein equations with conformal Killing $\eta$ fulfilling conditions (I)-(III) above. Then, and only then, $\Omega^{-2}g$ is a Fefferman-Graham ambient metric.
\end{teo}

To achieve such result, we identify a particularly convenient subclass of conformal rescalings, that we call \textit{geodesic Killing gauge}, in which the gradient of the conformal factor is geodesic and the conformal Killing becomes an actual Killing vector. Once $(\mc M,g,\Omega)$ is expressed in any such gauge, we construct a unique Rácz-Wald coordinate system, called \textit{adapted Rácz-Wald coordinates}, which allows us to relate the transverse expansion of the metric at the homothetic horizon to that at null infinity. Furthermore, we identify the remaining conformal freedom, consisting of arbitrary conformal transformations at the bifurcation surface. This residual freedom coincides precisely with the one in the original Fefferman-Graham construction. \\


Besides providing a conformal characterization of the ambient metric, the results in this paper establish an interesting link between the Fefferman-Graham expansion of the ambient metric across the homothetic horizon and at null infinity. This connection will play a significant role in our forthcoming paper \cite{Mio7} where we will address general existence and uniqueness of the asymptotic conformal equations from data at null infinity. As will be shown, the gravitational degrees of freedom of a general asymptotically flat spacetime appear at the same order as the free data in the Fefferman-Graham construction. Moreover, an intrinsic obstruction to the smoothness of $\scri$ is found to be closely linked to the Fefferman-Graham obstruction tensor at the cuts of $\scri$ being non-zero. This points to a deep relationship between the ambient metric and a geometric characterization of radiation, which we intend to investigate in future works.\\

This paper is organized as follows. In Section \ref{sec_ambientmetric} we review the Fefferman-Graham construction and its connection with the self-similar metrics studied in \cite{rodnianski}. In Section \ref{sec_compactification} we prove that the ambient metric admits a conformal extension and analyse its key properties. In Section \ref{sec_characterization} we identify a complete set of sufficient conditions that characterize the ambient metric from a conformal perspective. Finally, Appendix \ref{app} contains the explicit form of the quasi-Einstein equations in Rácz-Wald coordinates.

	\section*{Notation and conventions}
	
All manifolds in this paper are assumed to be connected and, depending on convenience, both index-free and abstract index notation are used to denote tensorial operations. Spacetime indices are denoted with Greek letters, indices on a hypersurface are written in lowercase Latin letters, and indices at cross-sections of a hypersurface are expressed in uppercase Latin letters. As usual, square brackets enclosing indices denote antisymmetrization and parenthesis are for symmetrization. By $\mc F(\mc M)$, $\X(\mc M)$ and $\X^{\star}(\mc M)$ we denote, respectively, the set of smooth functions, vector fields and one-forms on a manifold $\mc M$. The subset $\mc F^{\star}(\mc M)\subset\mc F(\mc M)$ consists of the nowhere vanishing functions on $\mc M$. A $(p,q)$-tensor refers to a tensor field $p$ times contravariant and $q$ times covariant. Given any pair of $(2,0)$ and $(0,2)$ tensors $A^{ab}$ and $B_{cd}$ we denote $\tr_A \bm B \d A^{ab}B_{ab}$.

	\section{Review of the ambient metric}
	\label{sec_ambientmetric}
	
In this section, we summarize the main aspects of the ambient metric construction developed in \cite{fefferman2012ambient}. The strategy we present differs slightly from the original one in order to simplify the exposition. The basic idea is to associate to each conformal structure of signature $(p,q)$ another ambient manifold of signature $(p+1,q+1)$, as we now review.\\

Let $(S,[h])$ be an $\mf n$-dimensional smooth conformal structure of signature $(p,q)$. Consider the product manifold $\real^+\times S$ and denote by $t$ the coordinate in its first factor. Then, after picking up a representative $h\in[h]$ one introduces the vector field $T\d t\partial_t$ and the degenerate bilinear form $\bm{\sigma} \d t^2 \pi^{\star}h$, where $\pi:\real^+\times S\to S$ is the projection onto the second factor. Note that $\lie_T \bm{\sigma} = 2\bm{\sigma}$. \\

Now consider the space $\real\times\real^+\times S$, and denote the coordinate in the first factor by $\rho$. The idea is to extend $T$ trivially off $\real^+\times S$, and to construct a smooth metric $g$ of signature $(p+1,q+1)$ in a neighbourhood of $\rho=0$, the so-called \textit{ambient metric}, such that (i) $T$ is a homothety of $g$, (ii) the pullback of $g$ into $\real^+\times S$ is $\bm{\sigma}$ and (iii) $g$ is Ricci-flat to infinite order at $\rho=0$. As shown in \cite{fefferman2012ambient}, it is sufficient to consider ambient metrics written in \textit{normal form}, i.e. such that at each point $(\rho=0,t,x)\in \{0\}\times \real^+\times S$ the metric is of the form $$g = 2t dt d\rho +\bm{\sigma},$$ and the vector $\partial_{\rho}$ is geodesic. The reason is that every ambient metric can be diffeomorphically mapped into one that is written in normal form. Taking all these considerations into account, the line element in a neighbourhood of $\rho=0$ can be written as 
\begin{equation}
	\label{ambientmetric}
	g = 2 a dt^2 + 2 t dt d\rho + 2 t b_A dt dx^A + t^2 \mu,
\end{equation}
where $a(\rho,x^A)$, $b_A(\rho,x^A)$ and $\mu(\rho,x^A)$ are to be determined as a series in $\rho$ by imposing Ricci flatness order by order at $\rho=0$. We use the notation $\mu^{(m)}$, $a^{(m)}$, $b^{(m)}$ to denote the $m$-th term of the expansion, i.e. $\mu^{(m)}\d \partial_{\rho}^{(m)}\mu |_{\rho=0}$, etc. By construction, one has $a^{(0)}=0$, $b^{(0)}=0$ and $\mu^{(0)}= h$.\\

The idea to obtain the expansion $\{a^{(m)},b^{(m)},\mu^{(m)}\}_{m\ge 1}$ is as follows. First, the $(t,t)$ and $(t,A)$ components of the ambient Ricci tensor at $\rho=0$ are $$R_{tt} \st{\rho=0}{=}\dfrac{\mf n}{t^2}( 1-a^{(1)}),\qquad R_{tA} \st{\rho=0}{=} \dfrac{1}{2t}\big(2\nabla_A^h a^{(1)} - \mf n b_A^{(1)}\big),$$ where $\nabla^h$ is the Levi-Civita connection of $h$. It is clear that only the choice $a^{(1)}=1$ and $b_A^{(1)}=0$ makes $R_{tt}\st{\rho=0}{=}0$ and $R_{tA}\st{\rho=0}{=}0$. After inserting them into the $(A,B)$ components of the ambient Ricci tensor at $\rho=0$ one arrives at $$R_{AB}\st{\rho=0}{=} \left(1-\dfrac{\mf n}{2}\right)\mu_{AB}^{(1)} -\dfrac{R^{(h)}}{2(\mf n-1)} h_{AB} + R^{(h)}_{AB},$$ where $R^{(h)}_{AB}$ is the Ricci tensor of $h$. Whenever $\mf n >2$, one can uniquely choose $\mu^{(1)}_{AB}$ to make $R_{AB}\st{\rho=0}{=}0$, and when $\mf n=2$ equation $R_{AB}\st{\rho=0}{=}0$ is automatically fulfilled because in dimension two every metric satisfies $R_{AB}^{(h)}=\frac{R^{(h)}}{2}h_{AB}$. Now suppose that $a^{(k)}$, $b^{(k)}$, $\mu^{(k)}$ ($1\le k \le m$) have already been determined. The $(t,t)$ component of the ambient Ricci tensor then reads $$\partial_{\rho}^m R_{tt} \st{\rho=0}{=} \dfrac{2m-{\mf n}}{t^2} a^{(m+1)} + \mc{L}^{(m)},$$ where $\mc{L}^{(m)}$ gathers lower order terms, i.e. terms depending on $a^{(m)}$, $b^{(m)}$, $\mu^{(m)}$ and below. For every $m$ when $\mf n$ is odd, and for every $m< \frac{\mf n}{2}$ when $\mf n$ is even, one can uniquely choose $a^{(m+1)}$ to make $\partial_{\rho}^m R_{tt} \st{\rho=0}{=} 0$. Substituting it into the $(t,A)$ components ($\mc{L}_A^{(m)}$ are again lower order terms), $$\partial_{\rho}^m R_{tA} \st{\rho=0}{=} \dfrac{2m-\mf n}{2t} b_A^{(m+1)} + \dfrac{1}{t}\nabla_A a^{(m+1)} + \mc{L}_A^{(m)},$$ allows one to determine $b_A^{(m+1)}$ so that $\partial_{\rho}^m R_{tA} \st{\rho=0}{=} 0$. Finally, the $(\rho,\rho)$ and $(A,B)$ components read (note the difference in the order of the transverse derivative in the left hand side)
\begin{equation}
	\label{Rrhorho}
	\partial^{m-1}_{\rho} R_{\rho\rho}\st{\rho=0}{=} -\dfrac{1}{2}h^{AB}\mu^{(m+1)}_{AB} + \mc{\wh L}^{(m)}
\end{equation}
and
\begin{equation}
	\label{partialmRij}
	\partial_{\rho}^{m}R_{AB} \st{\rho=0}{=}   \left(m+1-\dfrac{\mf n}{2}\right) \mu_{AB}^{(m+1)} - \dfrac{1}{2}\big(h^{CD}\mu^{(m+1)}_{CD}\big)h_{AB} + \nabla_{(A}b^{(m+1)}_{B)} + 2a^{(m+1)}L_{AB}+ \mc{L}^{(m)}_{AB},
\end{equation}
where $L_{AB}$ is the Schouten tensor\footnote{We recall the definition of this tensor in Section \ref{sec_compactification}, see \eqref{schouten}.} of $h$ and as before $\mc{\wh L}^{(m)}$ and $\mc{L}^{(m)}_{AB}$ gather all the lower order terms, i.e. the terms depending on $a^{(m)}$, $b^{(m)}$, $\mu^{(m)}$ and below. From \eqref{Rrhorho} one can always choose $h^{AB}\mu^{(m+1)}_{AB}$ so that $\partial^{m-1}_{\rho} R_{\rho\rho}\st{\rho=0}{=}0$ for every $m\ge 1$. The contracted Bianchi identity then implies that $h^{AB}\partial_{\rho}^{m}R_{AB} \st{\rho=0}{=} 0$, $\partial_{\rho}^{m-1}R_{\rho t} \st{\rho=0}{=} 0$ and $\partial_{\rho}^{m-1}R_{\rho A} \st{\rho=0}{=} 0$ are automatically fulfilled. As a consequence, the only non-trivial equation corresponds to the trace-free part of \eqref{partialmRij} w.r.t. $h$. Whenever $m< \mf n/2 -1$ (for $\mf n$ even) and for all $m$ (for $\mf n$ odd), one can uniquely choose $(\mu_{AB}^{(m+1)})^{TF}$ to make $\big(\partial_{\rho}^{m}R_{AB}\big)^{TF} \st{\rho=0}{=} 0$ (``TF'' denotes the trace-free part w.r.t. $h$). So, for $\mf n$ odd, the Einstein equations determine the full transverse expansion of the metric, $\{a^{(m)},b^{(m)},\mu^{(m)}\}_{m\ge 1}$, while for $\mf n$ even the expansion is determined up to and including order $\{a^{(\mf n/2)},b_A^{(\mf n/2)},\mu_{AB}^{(\mf n/2-1)}\}$, because only the trace of the coefficient $\mu^{(\mf n/2)}_{AB}$ can be determined from the equations. Both in the odd and even cases, $a=\rho$ and $b=0$ up to the order they are determined, i.e. $a^{(m)}=\delta^m_1$ and $b_A^{(m)}=0$ for all $m$ when $\mf n$ is odd, and for all $m=0,...,\mf n/2$ when $\mf n$ is even. In general, the metrics constructed in this way are called \textit{ambient metrics}.\\

As already mentioned, for $\mf n$ even the determination of $\mu^{(\mf n/2)}$ is obstructed due to the vanishing of the coefficient in front of $\mu^{(\mf n/2)}_{AB}$ in \eqref{partialmRij} when $m=\frac{\mf n}{2}-1$. Therefore, this equation does not fix the trace-free part of the term $\mu_{AB}^{(\mf n/2)}$, which can therefore be freely specified in the form of a $(0,2)$ symmetric, traceless tensor $\Psi_{AB}$. In addition, $\big(\partial_{\rho}^{\mf n/2-1}R_{AB}\big)^{TF}\st{\rho = 0}{=} 0$ holds if and only if $\big(\mc{L}^{(\mf n/2-1)}_{AB}\big)^{TF}$ vanishes identically. This tensor (which in general does not vanish) only depends on the initial metric $h$ and it is called \textit{obstruction tensor} $\mc{O}_{AB}$. It has the following properties (indices $A,B,...$ are raised and lowered with the metric $h$): (i) it is traceless $\mc{O}_A{}^A = 0$, (ii) divergence-free $\nabla^h_{B}\mc{O}_{A}{}^B = 0$, (iii) conformally covariant and (iv) it vanishes if $h$ is Einstein (but not only). As an example let us study the first non-trivial obstruction tensor, i.e. the one appearing for $\mf n=4$\footnote{As already discussed, for $\mf n=2$ equation $R_{AB}\st{\rho=0}{=}0$ holds automatically, and hence there is no obstruction tensor when $\mf n=2$.}. Equation \eqref{partialmRij} for $m=1$ is $$\partial_{\rho}R_{AB} \st{\rho = 0}{=} \left(2-\dfrac{\mf n}{2}\right)\mu^{(2)}_{AB} + 2 B_{AB}-2(\mf n-4) L_{A}{}^C L_{BC},$$ where $B$ is the Bach tensor of $h$. For $\mf n\neq 4$, this equation determines the coefficient $\mu^{(2)}$, but for $\mf n=4$ the trace-free part of $\mu^{(2)}$ remains undetermined and the ambient metric can only be Ricci flat provided the Bach tensor of $h$ (which is the obstruction tensor in dimension 4) vanishes. Thus, for $\mf n\ge 4$ even, it is not possible in general to have an ambient metric that is simultaneously more than $\mf n/2-1$ times differentiable and satisfy $R_{\alpha\beta}\st{\rho=0}{=}0$ to infinite order. If one insists on forcing $R_{\alpha\beta}\st{\rho=0}{=}0$ to infinite order (or, more strongly, in a neighbourhood of the homothetic horizon) then one must include logarithmic terms into the expansion that compensate the presence of the obstruction tensor, but at the same time spoil the smoothness of the ambient metric at the horizon $\rho=0$.\\

When $\mf n$ is odd, there is no obstruction tensor, but one can naturally consider solutions with expansions involving half-integral powers of $\rho$ which also have an indeterminacy at order $\mf n/2$ and that satisfy $R_{\alpha\beta}\st{\rho=0}{=}0$ to infinite order. These ambient metrics are, in general, of class $\mc C^{\lfloor \mf n/2\rfloor -1}$ at $\rho=0$. To sum up, ambient metrics satisfying $R_{\alpha\beta}\st{\rho=0}{=}0$ to infinite order present the following differentiability: for $\mf n=2$ they are smooth; when $\mf n\ge 4$ is even they are, generically, no more than $\mf n/2-1$ times differentiable; and for $\mf n\ge 3$ odd they are smooth provided there are no half-integer powers, or $\lfloor \mf n/2\rfloor-1$ times differentiable otherwise.\\




Given analytic data $(h,\Psi)$, standard convergence techniques establish existence of the ambient metric \cite{baouendi1976singular}, even in the presence of a non-vanishing obstruction tensor \cite{kichenassamy2004conjecture}. In odd dimensions, the works \cite{anderson2005existence,anderson2005asymptotically,kaminski2023well} show existence of ambient metrics beyond the analytic setting. In a remarkable breakthrough, the authors of \cite{rodnianski} study a class of metrics, called \textit{self-similar}, defined as metrics admitting a homothetic horizon and solving \textit{exactly} the Einstein equations. They work in double null coordinates in which the metric takes the form
\begin{equation}
	\label{metricdnc}
	g = -2\Theta^2 d\bar ud \bar v +\slashed{g}_{AB}\big(dx^A-q^Ad\bar v\big)\big(dx^B-q^Bd\bar v\big),
\end{equation}
 where $\Theta$, $q$ and $\slashed{g}$ are a function, a one-form and a Riemannian metric on the surfaces $S_{\bar u,\bar v}$, respectively. The homothetic vector in these coordinates is $K=\bar v\partial_{\bar v}+\bar u\partial_{\bar u}$ and the homothetic horizon is at $\{\bar u=0\}$. Their strategy is to define characteristic initial data on $\{\bar u=0\}\cup \{\bar v=-1\}$ to construct self-similar solutions after applying a suitable scaling limit. On $\{\bar u=0\}$ they specify the lapse $\Theta$, the shift $q$ and the conformal class subject to the normalization conditions $\Theta|_{\bar u=0}=1$ and $q|_{\bar u=0}=0$ (according to \cite{rodnianski} the condition on $\Theta$ is not very relevant, but the one on $q$ is essential to guarantee the regularity of $\Theta$). In principle, there are many ways to prescribe data on the hypersurface $\{\bar v=-1\}$, but after the limiting process used by the authors to construct self-similar solutions they show that by prescribing solely a Riemannian metric $h$ and symmetric traceless $(0,2)$-tensor $\Psi$, there exists a unique self-similar, \textit{regular} solution such that (i) $\slashed{g}|_{S_{0,-1}} = h$, (ii) $\Psi$ agrees with the trace-free part of the coefficient $v^{\mf n/2}$ of $\slashed{g}$ at $\bar u=0$ and (iii) $\slashed{g}$ satisfies the normalization conditions $\Theta|_{\bar u=0}=1$ and $q|_{\bar u=0}=0$.\\
 
The notion of regularity can be understood as follows. For $\mf n=2$, $g$ is regular provided it is smooth. When $\mf n\ge 3$ is odd, $g$ is regular provided it is smooth everywhere except at $\bar u=0$, and there exists smooth tensors $\{g^{(i)}_{\alpha\beta}\}_{i=0}^{\frac{\mf n-1}{2}}$ and $\wt{g}_{\alpha\beta}$ such that $$g_{\alpha\beta} = \sum_{i=0}^{\frac{\mf n-1}{2}} g_{\alpha\beta}^{(i)} \bar{u}^i +  \wt{g}_{\alpha\beta}\bar{u}^{\frac{\mf n}{2}} + O\big(\bar{u}^{\frac{\mf n+1}{2}}\big).$$ Finally, when $\mf n\ge 4$ is even, $g$ is regular if it is smooth everywhere except at $\bar u=0$, and there exists smooth tensors $\{g^{(i)}_{\alpha\beta}\}_{i=0}^{\frac{\mf n}{2}}$ and $\wt{g}_{\alpha\beta}$ such that $$g_{\alpha\beta} = \sum_{i=0}^{\frac{\mf n}{2}-1} g_{\alpha\beta}^{(i)} \bar{u}^i + g_{\alpha\beta}^{(\mf n/2)}\bar{u}^{\mf n/2} + \wt{g}_{\alpha\beta} \bar{u}^{\mf n/2} \log\bar u + O\big(\bar{u}^{\frac{\mf n}{2}+1} \log(\bar u)\big).$$ Observe that this definition of regularity is in complete agreement with the differentiability for Fefferman-Graham ambient metrics discussed above.\\

One natural question is whether an \textit{exact ambient metric} (i.e. one solving $R_{\alpha\beta}=0$ in a neighbourhood of the horizon) is self-similar in the sense of \cite{rodnianski} and vice-versa. There is one specific situation where exact ambient metrics and self-similar metrics are in one-to-one correspondence, namely when the ambient metric is \textit{straight}.

\subsection{Straightness}

One says that an ambient metric is straight provided the homothety $T$ satisfies $d\bm T=0$. Proposition 3.4 of \cite{fefferman2012ambient} shows that this is equivalent to $\bm T$ being exact in a full neighbourhood of the horizon, and also equivalent to $a=1$ and $b_A=0$. In fact, the metric \eqref{ambientmetric} with $a=1$ and $b_A=0$ satisfies $R_{tt}=R_{t\rho}=R_{tA}=0$ exactly. As we already discussed, one of the results in \cite{fefferman2012ambient} is that every ambient metric is straight up to the order determined by the initial metric $h$, i.e. to infinite order for $\mf n$ odd (and no half integers are allowed) and up to and including order $\mf n/2$ when $\mf n$ is even. The question of whether the metric is straight to infinite order depends on the free data $\Psi$. When $\mf n$ is even, the $(\mf n/2-1)$ $\rho$-derivative of the $(\rho,A)$ components of the Einstein equations read 
\begin{equation}
	\label{equationD}
	\partial_{\rho}^{\mf n/2-1}R_{\rho A} \st{\rho=0}{=} \dfrac{1}{2} b_A^{(\mf n/2+1)} + \nabla^B (\mu^{(\mf n/2)})^{TF}_{AB} - D_A,
\end{equation} 
where $D_A$ is a one-form that depends only on $h$. Hence, the condition $\nabla^B \Psi_{AB} = D_A$ is necessary for the ambient metric to be simultaneously Ricci flat and straight to infinite order. As proven in \cite[Thm. 3.10]{fefferman2012ambient}, it is also sufficient. When $\mf n$ is odd, the ambient metric is straight to infinite order if and only if $\nabla_B \Psi^{AB} = 0$ (see \cite[Thm. 3.9]{fefferman2012ambient}). One can combine these two conditions into a single one by just writing $\nabla_B \Psi^{AB} = D^A$, letting $D_A$ be identically zero when $\mf n$ is odd. \\


In the context of self-similar metrics, Proposition B.7 of \cite{rodnianski} shows that whenever the free data $\Psi$ satisfies $\nabla_B \Psi^{AB} = D^A$, the metric \eqref{metricdnc} has vanishing torsion one-form $\zeta$ and scalar $\omega$\footnote{In double null coordinates, the torsion $\zeta$ and the function $\omega$ are defined by $\zeta_A \d \dfrac{1}{2}g(\nabla_A e_a,e_3)$ and $\omega\d -\dfrac{1}{4}g(\nabla_4 e_3,e_4)$, where $e_3\d \Theta^{-1}\partial_{\bar v}$ and $e_3\d \Theta^{-1}(\partial_{\bar u}+q^A\partial_A)$ are null vectors satisfying $g(e_3,e_4)=-2$.}. These are given in terms of $\Theta$ and $q$ by $\omega = -\dfrac{1}{2}\nabla_4(\log\Theta)$ and $\zeta_A=-\dfrac{1}{4}\Theta^{-1} e_4(q^B)\slashed{g}_{AB}$. Thus, the self-similar solutions that satisfy $\nabla^B \Psi_{AB} = D_A$ also have $\Theta=1$ and $q=0$ everywhere. The change of coordinates $t = \bar v$ and $\rho = \bar u \bar v^{-1}$ in \eqref{metricdnc} leads to $$g = -2\rho dt^2 - 2t dt d\rho +\slashed{g}_{AB} dx^A dx^B,$$ which after identifying $\slashed{g} = t^2\mu$ happens to be a straight, exact ambient metric. In the rest of the paper we will focus on straight metrics, so we will not make a distinction between self-similar metrics and exact ambient metrics. One can rephrase Theorem 1.3 of \cite{rodnianski} in terms of straight, exact ambient metrics as follows.

\begin{teo}[\cite{rodnianski}]
	\label{teo_rodnianski}
Let $h$ be a Riemannian metric and $\Psi$ a symmetric, traceless tensor field on $S$ that satisfies $\nabla^B \Psi_{AB} = D_A$, where $D_A$ is an explicit one-form that only depends on $h$ when $\mf n$ is even, and identically zero for $\mf n$ odd. Then, there exists a unique straight, exact and regular ambient metric such that $g|_{S} = h$ and $\Psi$ agrees with the trace-free part of the coefficient $\bar u^{\mf n/2}$ of $g$ at the horizon.
\end{teo}

	\section{Conformal completion and null infinity}
	\label{sec_compactification}
	
	In this section we show that the ambient metric admits a conformal completion with a null conformal infinity. Before proving this we first recall the key aspects of the conformal Einstein equations. For details see \cite{friedrich2002conformal,frauendiener2004conformal,friedrich2015geometric}. We start by fixing some notation. Given a semi-Riemannian manifold $(M,g)$ of dimension $d\ge 3$ the Schouten tensor is defined by 
	\begin{equation}
		\label{schouten}
		\sch_g \d \dfrac{1}{d-2}\left(\ric_g -\dfrac{\scal_g}{2(d-1)} g\right),
	\end{equation}
	where $\ric_g$ and $\scal_g$ are the Ricci tensor and scalar, respectively. Reciprocally, the expression of the Ricci tensor in terms of the Schouten is given by
	\begin{equation}
R_{\alpha\beta} = (d-2)L_{\alpha\beta} + \dfrac{\scal_g}{2(d-1)}g_{\alpha\beta}	= (d-2) L_{\alpha\beta} + L g_{\alpha\beta},
	\end{equation}
where as in Section \ref{sec_ambientmetric} we employ the symbols $L_{\alpha\beta}$ and $L$ for the Schouten tensor and its trace in index notation, and we used that $L=\frac{\scal_g}{2(d-1)}$. From the transformation of the Ricci tensor under a conformal rescaling $g=\omega^m\wt{g}$, namely $${R}_{\alpha\beta} = \wt R_{\alpha\beta} - \dfrac{m(d-2)}{2}\left(\dfrac{\wt\nabla_{\alpha}\wt\nabla_{\beta}\omega}{\omega} -\dfrac{m+2}{2\omega^2}\wt\nabla_{\alpha}\omega\wt\nabla_{\beta}\omega\right)-\dfrac{m}{2}\left(\dfrac{\square_{\wt g} \omega}{\omega} +\dfrac{md-2(m+1)}{2\omega^2}|\wt\nabla\omega|^2_{\wt g}\right)\wt g_{\alpha\beta},$$ it follows that
	\begin{equation}
		\label{transshouten}
L^{tf}_{\alpha\beta} = \wt L^{tf}_{\alpha\beta} - \dfrac{m}{2}\left(\dfrac{\wt\nabla_{\alpha}\wt\nabla_{\beta}\omega}{\omega} -\dfrac{m+2}{2\omega^2}\wt\nabla_{\alpha}\omega\wt\nabla_{\beta}\omega\right)^{tf},
	\end{equation}
	where in this case ``tf'' denotes the trace-free part w.r.t. $g$ (or w.r.t. $\wt g$, since ``tf'' is a conformally invariant operation). Let $(M,[g])$ be a $d$-dimensional conformal structure. For each $g\in [g]$ one constructs the differential operator 
	\begin{equation}
		\label{operatorA}
		A_g(f)\d \big(\hess_g f + f\sch_g\big)^{tf},\qquad f\in\mc{F}(M).
	\end{equation}
	From the conformal transformation law of $\nabla$ and $\sch_g$ one can check that
	\begin{equation}
		\label{confproperty}
		A_{\omega^2 g}(\omega f) = \omega A_g(f).
	\end{equation}
	Recall that (in dimension $d\ge 3$) $g$ is an Einstein metric if and only if $\sch_g^{tf}=0$. Then, putting $f=1$ in \eqref{operatorA} and $\omega=\Omega$ in \eqref{operatorA}-\eqref{confproperty} one gets $$\Omega^{-2}g\in[g] \quad \mbox{is Einstein} \qquad \Longleftrightarrow \qquad \hess_g^{tf}\Omega + \Omega \sch_g^{tf}=0.$$ The previous equation can be written equivalently in terms of the scalar $\mf s\d d^{-1}\big(\square_g \Omega + \Omega L\big)$ as 
	\begin{equation}
		\label{QEequation}
		\hess_g \Omega + \Omega \sch_g -\mf s g = 0.
	\end{equation}
	A 3-tuple $(M,g,\Omega)$ is a (vacuum) quasi-Einstein manifold provided that \eqref{QEequation} is satisfied. Property \eqref{confproperty} guarantees that if $(M,g,\Omega)$ is quasi-Einstein and $\omega\in\mc{F}^{\star}(M)$, then $(M,\omega^2g,\omega\Omega)$ is also quasi-Einstein. One important consequence of this definition is that the quantity 
	\begin{equation}
		\label{lambda}
		\lambda \d 2\mf s\Omega -|\nabla\Omega|_g^2
	\end{equation}
	is constant (and conformally invariant). Ignoring an irrelevant numerical factor, the constant $\lambda$ corresponds to the cosmological constant associated to $\Omega^{-2}g$, which is the Einstein representative of $(M,g,\Omega)$.

	\subsection{Conformal completion of the ambient metric}


Let us consider a straight ambient metric
\begin{equation}
	\label{ambientstraight}
	g= 2\rho dt^2 + 2t dt d\rho + t^2\mu
\end{equation}
and the change of coordinates $\{t,\rho\}\mapsto \{t, u\d \rho t\}$, under which \eqref{ambientstraight} takes the form 
\begin{equation}
	\label{ambientmetric2}
	g = 2dt du + t^2 \mu,
\end{equation}
where now $\mu$ is a series in powers of $\frac{u}{t}$. In these coordinates the homothetic horizon ($\rho=0$) is placed at $u=0$, and the infinity is reached when $t\sto \infty$. Introducing the coordinate $v \d t^{-1}$ the metric \eqref{ambientmetric2} becomes 
\begin{equation}
	\label{ambientmetriccomp}
g = v^{-2}\big(-2du dv + \mu \big),
\end{equation}
where now $\mu$ is a series in powers of $uv$. Then, the metric 
\begin{equation}
	\label{ambientmetriccomp2}
	\wt{g} \d v^2 g = -2du dv + \mu
\end{equation}
 is extendible beyond $\scri\d\{v = 0\}$, the (null) conformal infinity of the ambient metric. Note that if $g$ is Ricci flat, then $(\wt g,v)$ satisfies \eqref{QEequation}. Following the notation of Section \ref{sec_ambientmetric}, for $\mf n$ odd the tensor $\mu$ is given as a formal series by $$\mu = \sum_{k=0}^{\frac{\mf n-1}{2}} \dfrac{1}{k!}\mu^{(k)} (uv)^k + \wt{\mu} (uv)^{\frac{\mf n}{2}} + O\big((uv)^{\frac{\mf n+1}{2}}\big),$$ while for $\mf n\ge 4$ even $\mu$ is given by $$\mu = \sum_{k=0}^{\frac{\mf n}{2}-1} \dfrac{1}{k!}\mu^{(k)} (uv)^k + \mu^{(\mf n/2)} (uv)^{\mf n/2}+ \wt{\mu} (uv)^{\frac{\mf n}{2}}\log(uv) + O\big((uv)^{\frac{\mf n}{2}+1}\log(uv)\big).$$ As already said, the case $\mf n=2$ is special because no obstruction tensor appears and the ambient metric is always smooth.\\
 
 Note that in these coordinates the vector $T$ is given by $T = u\partial_u - v\partial_{v}$. It is straightforward to check that $T$ is a Killing vector w.r.t $\wt g$ and that the set $\{u=0\}\cup \{v = 0\}$ is a (non-degenerate) bifurcate Killing horizon with bifurcation surface $\{u=v=0\}$. Also note that the straightness condition implies that the one-form $\wt{\bm T}\d \wt{g}(T,\cdot)$ is integrable in the conformal manifold $(\wt{M}\d M\cup\scri,\wt g,\Omega)$, i.e. $\wt{\bm T}\wedge d\wt{\bm T}=0$. Being part of a Killing horizon, it follows that $\scri$ is a totally geodesic null hypersurface with first fundamental form $h$. Moreover, the transverse derivatives of $\wt g$ at $\{v=0\}$ are $$\partial^{m}_{v} \wt{g} \st{\scri}{=} \mu^{(m)} u^m, \qquad 1\le m\le \lceil \mf n/2-1\rceil,$$ while the transverse derivatives along $s\d uv$ at $s=0$ are given by\footnote{We use $\partial_s$ as the derivative w.r.t. the product $uv$ for functions that only depend on $(uv,x^A)$.}
 \begin{equation}
 	\label{partials}
 	\partial^{m}_{s} \wt{g} \st{s=0}{=} \mu^{(m)},\qquad 1\le m\le \lceil \mf n/2-1\rceil.
 \end{equation} 
 This means that the free data $\Psi_{AB}$ agrees with the trace-free part of the coefficient in $s^{\mf n/2}$.\\

Particularizing equation \eqref{lambda} to $\lambda=0$ and taking into account that $|\wt\nabla\Omega|^2_{\wt g}=|\wt\nabla v|^2_{\wt g} = 0$ it follows that $\mf s=0$, so $\mc{Q}_{\alpha\beta}\d \wt{\nabla}_{\alpha}\wt{\nabla}_{\beta}\Omega + \Omega \wt{L}_{\alpha\beta}=\mc{Q}_{\alpha\beta}^{tf}$ and thus the quasi-Einstein equations can be written as $\mc{Q}_{\alpha\beta} = 0$. In other words, if $g$ is exactly Ricci flat, then $\wt g$ satisfies $\mc Q_{\alpha\beta}=0$. Let us consider an even dimensional ambient metric constructed by solving $\partial_{\rho}^{(m)}R_{\alpha\beta}\st{\rho=0}{=} 0$ up to an including order $m=\mf n/2-2$. As we discussed before, the determination of the next coefficient of the expansion is (in general) obstructed, in the sense that $(\partial^{(\mf n/2-1)}_{\rho} R_{AB})^{TF} \st{\rho=0}{=} \mc{O}_{AB}\neq 0$, where $\mc{O}_{AB}$ is the obstruction tensor of $h=\mu^{(0)}$. This obstruction can also be detected at $\scri$, as we explain next.\\

First observe that $\partial_{\rho}^{m}R_{\alpha\beta}\st{\rho=0}{=} 0 \quad \forall m\le \mf n/2-2$ implies $\partial_{\rho}^{m}L_{\alpha\beta}\st{\rho=0}{=} 0$ and $\partial_{\rho}^{m}R\st{\rho=0}{=}\partial_{\rho}^{m}L\st{\rho=0}{=}0\quad\forall m\le \mf n/2-2$ ($R$ and $L$ are the Ricci and Schouten scalars of $g$, respectively), so (cf. \eqref{schouten}) $$\mc{O}_{AB} = (\partial^{\mf n/2-1}_{\rho} R_{AB}|_{\rho=0})^{TF} = \left(\mf n \partial_{\rho}^{\mf n/2-1} L_{AB}|_{\rho=0}+\dfrac{\partial_{\rho}^{\mf n/2-1}R|_{\rho=0}}{2(\mf n+1)} h_{AB}\right)^{TF} ,$$ and therefore $$\mc{O}_{AB}= \mf{n}(\partial_{\rho}^{\mf n/2-1} L_{AB}|_{\rho=0})^{TF}.$$

Recalling that $L_{\alpha\beta}^{tf} = \wt{L}^{tf}_{\alpha\beta} + \Omega^{-1}\big(\hess_{\wt g}\Omega\big)^{tf}_{\alpha\beta} = \Omega^{-1}\mc{Q}_{\alpha\beta}$ (see \eqref{transshouten}) it follows that $$L_{\alpha\beta}-\dfrac{L}{\mf n+2} g_{\alpha\beta} = \Omega^{-1}\mc{Q}_{\alpha\beta}\qquad\Longrightarrow\qquad L_{AB}-\dfrac{L}{\mf n+2} t^2\mu_{AB} = \Omega^{-1}\mc{Q}_{AB},$$ and then (note that $\partial_{\rho}\Omega = \partial_{\rho} t^{-1} = 0$) $$\partial_{\rho}^{\mf n/2-1}L_{AB}|_{\rho=0} - \dfrac{\partial_{\rho}^{\mf n/2-1}L|_{\rho=0}}{\mf n+2}t^2 h_{AB} = \Omega^{-1}\partial_{\rho}^{\mf n/2-1}\mc{Q}_{AB}|_{\rho=0}.$$ Under the transformation $\{t=v^{-1},\rho=uv\}$ this expression becomes (note that $\partial_{\rho}=v^{-1}\partial_u$) $$(\partial_{\rho}^{\mf n/2-1}L_{AB}|_{\rho=0})^{TF} = v^{-\mf n/2}(\partial_{u}^{\mf n/2-1}\mc{Q}_{AB}|_{u=0})^{TF} \qquad\Longrightarrow\qquad \mc{O}_{AB} = \mf{n}v^{-\mf n/2}(\partial_u^{\mf n/2-1}\mc{Q}_{AB}|_{u=0})^{TF}.$$ 

We are interested in obtaining an expression relating the obstruction tensor with transverse derivatives of the tensor $\mc Q$ at null infinity, i.e. derivatives of $\mc Q$ w.r.t. $\partial_v$ at $v=0$. In other words, we want to ``interchange'' the roles of $u$ and $v$ in the formula for $\mc{O}_{AB}$ we have just obtained. In order to do that, we exploit the fact that $uv=0$ is a bifurcate Killing horizon. Indeed, since $\mc{Q} = \hess_{\wt g}v + v\sch_{\wt g}$ and $T=u\partial_u-v\partial_v$ is a Killing vector of $\wt g$, one has $\lie_{T}\mc{Q} = -\mc{Q}$ and hence $T(\mc{Q}_{AB})=-\mc{Q}_{AB}$. This implies that $\mc{Q}_{AB}$ is of the form $\mc{Q}_{AB} = v T_{AB}(uv,x^C)$, and hence $$\mc{O}_{AB} = \mf{n}v^{-\mf n/2+1}\big((\partial_u^{\mf n/2-1}T_{AB})|_{u=0}\big)^{TF}= \mf n \big(T^{(\mf n/2-1)}_{AB}|_{uv=0}\big)^{TF},$$ where $T^{(m)}$ denotes the $m$-th derivative w.r.t. $uv$ and we used that $T_{AB}(uv,x^C)|_{u=0}= T_{AB}(uv,x^C)|_{uv=0}$. Recall that $\partial_u^{k}=v^k\partial_s^{k}$ and $\partial_v^{k}=u^k\partial_s^{k}$ for every $k$, and note also that $$k!\big(T^{(k)}_{AB}|_{uv=0}\big)^{TF} = \big(\partial_u^{k}(u^k T^{(k)}_{AB})|_{v=0}\big)^{TF} =  \big(\partial_u^{k}\big(\partial_v^{k} T_{AB}\big)|_{v=0}\big)^{TF},$$ where we have used that $\partial_u^{k}(u^k T^{(k)}_{AB})|_{v=0}=k!T^{(k)}_{AB}|_{v=0}$. Note also that the horizon is totally geodesic and then $\partial_u$ and $\partial_v$ commute with $TF$. Then, 

$$\left(\dfrac{\mf n}{2}-1\right)!\, \mc{O}_{AB}=\mf n \partial_u^{\mf n/2-1}\big( (\partial_{v}^{\mf n/2-1}T_{AB})|_{v=0}\big)^{TF},$$ and since $\big(\partial_v^{\mf n/2}(vT_{AB})\big)|_{v=0} = \frac{\mf n}{2}\big(\partial_v^{\mf n/2-1} T_{AB}\big)|_{v=0}$, we conclude that $$\left(\dfrac{\mf n}{2}-1\right)!\, \mc{O}_{AB}= 2 \partial_u^{\mf n/2-1}\big( (\partial_{v}^{\mf n/2}\mc{Q}_{AB})|_{v=0}\big)^{TF} .$$

This shows that the presence of a non-vanishing obstruction tensor can also be detected from the conformal infinity, as it makes the $\mf n/2$-transverse derivative of the tensor $\mc{Q}$ at $\scri$ to be different from zero. 

\section{Conformal characterization of the ambient metric}
\label{sec_characterization}

In this section we characterize the ambient metric from a conformal viewpoint. More specifically, we want to find a set of conformally covariant conditions on a conformal manifold $(\mc M,g,\Omega)$ that univocally lead to the ambient metric. As we studied in the previous section, the conformal completion of the ambient metric exhibits two key properties, namely the existence of a bifurcate Killing horizon where one of the horizons is $\scri$, and that the one-form $\wt{\bm T}$ is integrable. While the latter condition is already conformally invariant, the former is not. The obvious replacement is to ask for the existence of a bifurcate \textit{conformal} Killing horizon. Additionally, it is clear that one should impose some extra condition relating the conformal Killing field (that we now denote by $\eta$ to avoid confusion with the previous section) and the function $\Omega$. In the conformal completion of the previous section $\Omega=v$ and $T=u\partial_u - v\partial_v$, so $T(\Omega) = -v$. The natural replacement for this condition is to ask that $\eta(\Omega) = \big(\psi-1\big)\Omega$, where $\psi$ is the function such that $\lie_{\eta}g = 2\psi g$. That this condition is conformally invariant follows because under a conformal rescaling $g' \d \omega^2 g$ the function $\psi$ transforms by $\psi' = \psi + \eta\big(\log|\omega|\big)$, and hence $\eta(\omega\Omega) = \omega \eta(\Omega) + \omega\Omega \eta\big(\log|\omega|\big) = \big(\psi'-1)\omega\Omega$. The main result of this paper is that these three conditions fully characterize the Fefferman-Graham ambient metric (Theorem \ref{teorema}). We start by showing that the function $\psi$ necessarily vanishes at the bifurcation surface.
\begin{lema}
	\label{lemapsi=0}
	Let $\eta$ be a conformal Killing field and $\psi$ the function defined by $\lie_{\eta}g = 2\psi g$. Assume $\eta$ admits a bifurcation surface $\mc S$. Then, $\psi|_{\mc S}=0$.
	\begin{proof}
Let $V,W\in T\mc S$. Since $[\eta,V]\st{\mc S}{=} 0$ it follows $\nabla_V\eta\st{\mc S}{=} 0$, and then $$0\st{\mc S}{=}V^{\alpha}W^{\beta} \nabla_{(\alpha}\eta_{\beta)} = \psi g(V,W) .$$ Since $V,W$ are arbitrary it follows that $\psi\st{\mc S}{=}0$.
	\end{proof}
\end{lema}

Next we show that it is possible to fix the gauge such that $|\nabla\Omega|^2 = 0$. 

\begin{lema}
	\label{lemagaugeconf}
	Let $(\mc M,g,\Omega)$ be a conformal manifold with $\lambda=0$ and $\mc H$ a hypersurface transverse to $\scri$. Let $\omega_0$ be a non-vanishing function on $\mc H$. Then, there exists a unique $(\wt g=\omega^2g,\wt \Omega=\omega\Omega)$ in a neighbourhood of $\scri\cap\mc H$ such that $|\wt{\nabla}\wt\Omega|^2_{\wt g}=0$ and $\wt\Omega\st{\mc H}{=}\omega_0\Omega$. 
	\begin{proof}
Let $(M,g,\Omega)$ be a conformal manifold, $\omega>0$ a function, and define $\wt{g}\d \omega^2 g$, $\wt{\Omega}\d \omega\Omega$ and $F \d |\nabla\Omega|^2_g$. Then, $$\wt{F}\d |\wt{\nabla}\wt{\Omega}|_{\wt g}^2 = \Omega^2\nabla_{\alpha} f \nabla^{\alpha}f +2\Omega \nabla_{\alpha} f \nabla^{\alpha}\Omega + F,$$ where we defined $f\d \log\omega$. We want to show that the equation $\wt F=0$ admits a unique solution given the function $f$ in a hypersurface transverse to $\scri$. Since $F|_{\Omega = 0}=0$, the function $\Omega^{-1}F$ has a good limit at $\Omega=0$ and henceforth we can equivalently look for solutions to the following equation
	\begin{equation}
	\label{Fprima}
	\wt F' = \Omega\nabla_{\alpha} f \nabla^{\alpha}f +2 \nabla_{\alpha} f \nabla^{\alpha}\Omega + \Omega^{-1} F=0.
\end{equation}
In order to do that we use the method of characteristics (see \cite{evans2022partial}), which basically consists of rewriting the PDE as a first order system of ODEs for $f$ and its gradient $p^{\alpha}\d \nabla^{\alpha}f$ along the so-called characteristic curves, with the aim of solving it from data at a hypersurface $\mc H$. In our specific setup, since $\nabla^{\alpha}\Omega$ is tangent to $\scri$, it is also necessarily transverse to $\mc H$ (at least in a neighbourhood of $\mc H\cap \scri$). Then, we can complete any local coordinate system $\{x^a\}$ on $\mc H$ to a local coordinate system $\{x^0,x^a\}$ on $M$ by extending $\{x^a\}$ trivially along $\nabla\Omega$, and solving $\nabla\Omega(x^0)=1$, $x^0|_{\mc H}=0$ (then, $\nabla\Omega  = \partial_{x^0}$). Given a smooth function $f$ on ${\mc H}$, an initial condition for $p$ is called admissible provided that $p_a|_{\mc H} = df_a$ and $\wt{F}'(f,p|_{\mc H}) \st{\mc H}{=} 0$. Note that, in general, a covector $p|_{\mc H}$ satisfying these conditions may not exist or may not be unique. It is only when the problem is non-characteristic, i.e. $\xi^{\alpha} \mc{D}_{p_{\alpha}}\wt F'|_{\mc H} \neq 0$ for any non-vanishing vector $\xi$ transverse to $\mc H$, when a unique solution for $p$ exists. In our specific setup, choosing $\xi=\nabla\Omega=\partial_{x^0}$, equation \eqref{Fprima} becomes
\begin{equation}
\wt F' = \Omega p_{\alpha}p^{\alpha}  + 2 p_{0} + \Omega^{-1}F=0,
\end{equation} 
and then one can check that $\xi^{\alpha} \mc{D}_{p_{\alpha}}\wt F'|_{\mc H} = \mc{D}_{p_0}\wt{F}' = 2+O(\Omega)$, so $\mc{D}_{p_0}\wt{F}'\neq 0$ in a neighbourhood of $\mc H\cap \scri$. This proves that the problem is non-characteristic and therefore the equation $|\nabla\Omega|_g^2=0$ admits a unique solution given the function $\omega$ on a hypersurface transverse to $\scri$.
	\end{proof}
\end{lema}

Let $(M,g,\Omega)$ be a conformal manifold admitting a bifurcate conformal Killing horizon such that one of the horizons is $\scri$. Let $\eta$ be the conformal field and $\psi$ the function satisfying $\lie_{\eta} g = 2\psi g$. Assume also that $\eta(\Omega)=\big(\psi-1\big)\Omega$. Now we show that one can restrict further the gauge to set $\psi=0$. 

\begin{prop}
	\label{prop_GKG}
	Let $(\mc M,g,\Omega)$ be a conformal manifold admitting a bifurcate conformal Killing vector $\eta$ where one of the horizons is $\scri$ and $\eta(\Omega)=(\psi-1)\Omega$. Then, there exists a conformal gauge in which simultaneously $|\nabla\Omega|^2=0$ and $\lie_{\eta} g = 0$ in a neighbourhood of the bifurcation surface. Moreover, the remaining gauge freedom is a function $\omega(x^A)$.
	\begin{proof}
We choose the transverse hypersurface $\mc H$ of the previous proposition as the conformal Killing horizon transverse to $\scri$, so that $\eta$ is tangent to $\mc H$ and $\mc S\d \mc H\cap \scri$ is the bifurcation surface of $\eta$. The idea of the proof is to show that one can make $\psi$ to vanish at $\mc H$ by choosing the free function $\omega$ of the proposition to satisfy $\psi' = \psi + \eta(\log\omega) = 0$ at $\mc H$, and then to prove that condition $|\nabla\Omega|^2=0$ implies that $\psi$ vanishes everywhere. Recall by Lemma \ref{lemapsi=0} that $\psi=0$ at $\mc H\cap \scri$, so in order to prove that the equation $\psi + \eta(\log\omega) = 0$ along $\mc H$ admits a solution we must show that $\eta|_{\mc H}$ has a zero of order one at $\mc S$. We choose double null coordinates $\{v,u,x^A\}$ adapted to $\scri=\{v=0\}$ and $\mc H=\{u=0\}$, where the metric is given by $$g = 2G du dv + \slashed{g}_{AB}(dx^A+q^A dv)(dx^B+q^B dv).$$  Note that $n\d \partial_v - q^A\partial_{x^A}$ is a null generator of $\mc H$ and by construction $\eta|_{\mc H} = H n$ for some function $H$ defined on $\mc H$. We want to show that $H$ satisfies $H|_{v=0}=0$ and $\partial_v H|_{v=0}\neq 0$.\\

Consider the equation $\eta^{\alpha}\nabla_{\alpha}\Omega = (\psi-1)\Omega$. Applying $\nabla_{\beta}$ to both sides it follows that $$\nabla_{\beta}\eta^{\alpha} \nabla_{\alpha}\Omega + \eta^{\alpha}\nabla_{\beta}\nabla_{\alpha}\Omega = \Omega\nabla_{\beta}\psi + (\psi-1)\nabla_{\beta} \Omega.$$ Defining $F_{\alpha\beta} \d \nabla_{[\alpha}\eta_{\beta]}$ and inserting $\nabla_{\beta}\eta^{\alpha} = \psi\delta_{\beta}^{\alpha}+F_{\beta}{}^{\alpha}$, $$\psi\nabla_{\beta}\Omega + F_{\beta}{}^{\alpha} \nabla_{\alpha}\Omega  + \eta^{\alpha}\nabla_{\beta}\nabla_{\alpha}\Omega = \Omega\nabla_{\beta}\psi + (\psi-1)\nabla_{\beta} \Omega,$$ which evaluated at $\mc S$ gives $F_{\beta}{}^{\alpha} \nabla_{\alpha}\Omega = -\nabla_{\beta} \Omega$. Since $d\Omega = f dv$ on $v=0$ with $f\neq 0$ it follows that $F_{\beta}{}^v = -\delta_{\beta}^v$, and hence taking $\beta=v$ and using that $g^{v\alpha}=G^{-1}\delta^{\alpha}_u$ we get $F_{vu}\st{\mc S}{=}\partial_v\eta_{u}-\partial_u\eta_v\st{\mc S}{=} 2\partial_v\eta_{u} \st{\mc S}{=} - G\neq 0$, where we used $\partial_{\alpha}\eta_{\beta}+\partial_{\beta}\eta_{\alpha}\st{\mc S}{=}0$. Since $\bm\eta|_{\mc H} = g(\eta,\cdot)|_{\mc H}=H du$ we conclude $\partial_v H|_{\mc S}\neq 0$, so $\eta|_{\mc H}$ has a zero of order one at $\mc S$ and then equation $\psi + \eta(\log\omega) = 0$ admits a solution along $\mc H$.\\

Once we have guaranteed that (part of) the remaining freedom in Prop. \ref{lemagaugeconf} can be chosen so that $\psi=0$ on $\mc H$, it remains to show that $\psi$ actually vanishes everywhere. Using that $\lie_{\eta}(g^{\alpha\beta})=-2\psi g^{\alpha\beta}$,
\begin{align*}
	\lie_{\eta}|\nabla\Omega|^2 &= -2\psi|\nabla\Omega|^2 +2\nabla^{\alpha}\Omega \nabla_{\alpha}((\psi-1)\Omega)\\
	&=-2\psi |\nabla\Omega|^2 + 2\Omega\nabla^{\alpha}\Omega \nabla_{\alpha}\psi +2(\psi-1)|\nabla\Omega|^2\\
	&= 2\big(\Omega\nabla^{\alpha}\Omega \nabla_{\alpha}\psi-|\nabla\Omega|^2\big).
\end{align*}
Then, in a gauge in which $|\nabla\Omega|^2=0$ one has $\nabla_{\alpha}\Omega\nabla^{\alpha}\psi=0$, and since $\nabla_{\alpha}\Omega$ is transverse to $\mc H$, the condition $\psi\st{\mc H}{=}0$ extends to a neighbourhood of $\mc S$.
	\end{proof}
\end{prop}

An immediate corollary of this proposition is the following.
\begin{cor}
	\label{cor_f1f2}
	Let $f_1$ and $f_2$ be two functions satisfying $|\nabla f_1|^2 =|\nabla f_2|^2= 0$ and $\eta(f_1)=-f_1$, $\eta(f_2)=-f_2$. Assume $f_1 \st{\mc S}{=} f_2$. Then, $f_1=f_2$.
	\begin{proof}
In Proposition \ref{prop_GKG} it is shown that the solution to $|\nabla f|^2=0$ and $\eta(f)=-f$ is characterized by a free function $\omega$ on $\mc S$. Since both $f_1$ and $f_2$ agree on $\mc S$ the function $\omega$ is identically $1$, and then $f_1=f_2$ everywhere.
	\end{proof}
\end{cor}

In this conformal gauge, from now on called \textit{geodesic Killing gauge}, the metric $g$ admits a bifurcate Killing horizon, so it can be written in Rácz-Wald coordinates as \cite{racz1992extensions}
\begin{equation}
	\label{metricRW}
	\wt g = 2G du \big(dv + v\beta_A dx^A\big) + \mu_{AB} dx^A dx^B,
\end{equation}
 where $G$, $\bm{\beta}$ and $\mu$ are a function, a one-form and a metric on the codimension-two surfaces $S_{u,v}$ that depend only upon $\{s\d uv,x^A\}$. In these coordinates the Killing field is given by $\eta = u\partial_u - v\partial_v$. We note that in the geodesic Killing gauge, the vector field $\grad \Omega$ is geodesic (hence the name). This follows from $0 =\nabla_{\beta} (\nabla_{\alpha} \Omega \nabla^{\alpha} \Omega) = 2 \nabla^{\alpha} \Omega \nabla_{\alpha} \nabla_{\beta} \Omega =0$.\\
 
The Rácz–Wald (RW) construction allows for the freedom to choose any cross-section $\Sigma$ on $\{v=0\}$ (not intersecting the bifurcation surface $\mc S$). From this surface, the coordinate $u$ is uniquely defined so that $u|_{\mc S}=0$, $u|_{\Sigma}=1$, and $\partial_u$ is geodesic. Given any conformal factor $\Omega$ in a geodesic Killing gauge, we fix the RW coordinate $u$ as follows. Choose the family of geodesic curves $\gamma(\tau)$ on $\{v=0\}$ satisfying $\tau|_{\mc S}=0$ and $\dot\gamma|_{\mc S} = \grad\Omega|_{\mc S}$ (note that $\grad\Omega$ is tangential to the hypersurface $\{v=0\}$). We then choose $\Sigma\d \{\tau=1\}$ in the construction described above. Since $u|_{\mc S}=\tau|_{\mc S}=0$, $u|_{\Sigma}=\tau|_{\Sigma}=1$, and both $\partial_u$ and $\dot\gamma$ are geodesic, it follows that $u=\tau$, so $\grad\Omega \st{v=0}{=} \partial_u$. In particular, $d\Omega\st{\mc S}{=} G dv$. The resulting RW coordinates only admit an additional scaling freedom for the coordinate $v$ of the form $v = \bar{v} h(x^A)$. Next, we show that one can rescale the coordinate $v$ to set $\bm\beta = 0$ locally when $\eta$ is integrable.

\begin{lema}
	Assume that the Killing $\eta=u\partial_u-v\partial_v$ is integrable w.r.t the metric \eqref{metricRW}. Then $\bm\beta$ is closed, and hence locally exact.
	\begin{proof}
		It is easier to work away from $u=0$ or $v=0$ and define the one-form $\wt{\bm \eta}\d (Guv)^{-1}\wt{g}(\eta,\cdot) = \bm\beta-d\log\big|\frac{u}{v}\big|$. Now, condition $\bm \eta\wedge d\bm \eta=0$ is equivalent to $\wt{\bm \eta}\wedge d\wt{\bm \eta}=0$, and hence $$0=\left(\bm\beta-d\log\left|\dfrac{u}{v}\right|\right)\wedge  d\bm\beta =  \bm\beta \wedge d s\wedge\dot{\bm\beta}+\bm\beta\wedge \slashed{d}\bm\beta - d \log\left|\dfrac{u}{v}\right| \wedge ds\wedge\dot{\bm\beta}-d\log\left|\dfrac{u}{v}\right|\wedge \slashed{d}{\bm\beta},$$ where the dot denotes derivative w.r.t. $s$ and $\slashed{d}$ is the exterior derivative on the codimension two surfaces $S_{u,v}$. Since the four terms on the RHS are linearly independent, the last term shows that $\slashed{d}\bm\beta = 0$ (note that although the computation has been done away from $u=0$ or $v=0$, by continuity the result is valid everywhere).
	\end{proof}
\end{lema}

\begin{cor}
	\label{coro}
	Assume the Killing $\eta=u\partial_u-v\partial_v$ is integrable w.r.t the metric \eqref{metricRW}. Then, there exists a change of coordinates that respects the form of $\eta$ and such as the metric takes the form \eqref{metricRW} with $\bm\beta=0$, i.e.
	\begin{equation}
		\label{metricRWbeta=0}
		g = 2G du dv + \mu.
	\end{equation}
	\begin{proof}
		Since $\bm \eta$ is integrable, by the previous lemma it is locally exact, i.e. there exists a function $f$ such that $\bm\beta = df$. Moreover, since $\dot{\bm\beta}=0$ it follows that $f=f(x^A)$. Inserting this and $v =\bar v h(x^A)$ into \eqref{metricRW} yields $$g = 2G du \big(\bar v dh + hd\bar v + h\bar v df\big) + \mu.$$ By choosing $h$ such that $dh +hdf = 0$ and redefining $G$, the metric $g$ takes (locally) the form \eqref{metricRWbeta=0}. Note that this change keeps the same form of $\eta$, since $v\partial_v =h \bar{v}h^{-1}\partial_{\bar v}= \bar{v}\partial_{\bar v}$.
	\end{proof}
\end{cor}

Note that this change of $v$ does not affect the coordinate $u$ and that the remaining freedom is scaling $v$ by a non-zero constant. In these new coordinates, we still have $d\Omega \st{\mc S}{=} G dv$. This means that $\Omega = vF$, where $F$ satisfies $F|_{\mc S}=G|_{\mc S}$. Since $\eta(\Omega)=-\Omega$ and $\eta(v)=-v$ it follows that $\eta(F)=0$ and thus $F=F(uv,x^A)$. \\

Next, we define $\wh\Omega \d F^{-1}\Omega$ and $\wh g \d F^{-2} g$ (note that $\wh\Omega=v$). We write the metric as $\wh g=2\wh G du dv +\wh\mu$ and note that $\wh G=F^{-2}G$, and in particular $\wh G|_{\mc S}=F^{-1}|_{\mc S}$ (this is the key reason for our specific choice of RW coordinates above). Using the conformal covariance of the equations, showing that $(g,\Omega)$ is quasi-Einstein is equivalent to showing that $(\wh g,v)$ satisfy $\wh{\mc Q} \d \hess v + v\sch_{\wh g} - \mf{s}\wh g=0$ (note that a priori the conformal factor $\wh\Omega$ need not to satisfy $|\nabla\wh\Omega|^2_{\wh g}=0$, i.e. we have momentarily abandoned the geodesic Killing gauge and the term $\mf{s}\wh g$ appears in the conformal Killing equations). In Appendix \ref{app} we have computed the components of the tensor $\hess v + v\sch_{\wh g}$ for the metric \eqref{metricRWbeta=0}. In particular, we are interested in the equations $\mc{Q}_{vv}=0$ and $\mc{Q}_{vA}=0$. Since $\wh{g}_{vv}=\wh{g}_{vA}=0$, the $(v,v)$ and $(v,A)$ equations that $(\wh g,v)$ satisfy are obtained by simply replacing $G$ by $\wh G$ and $\mu$ by $\wh\mu$. In particular, equation $\mc{Q}_{vv}=0$ is $$2\big(s \tr_{\wh\mu}\dot{\wh\mu}-2\mf n   \big)\dfrac{\dot{\wh G}}{\wh G}+ s\big(|\dot{\wh\mu}|^2- 2\tr_{\wh\mu}\ddot{\wh\mu}\big)=0,$$ which proves $\dot{\wh G} = 0$ (at least in a neighbourhood of $s=0$), and equation $\mc{Q}_{vA}=0$ reads $$\big(s\tr_{\wh\mu}\dot{\wh\mu}-2\mf n  \big)\dfrac{\nabla_A \wh G}{\wh G} - 2s\big(\nabla_A\tr_{\wh\mu}\dot{\wh\mu} - (\div_{\wh\mu}\dot{\wh\mu})_A\big)=0,$$ which evaluated at $s=0$ implies $\wh G=A\in\real\setminus\{0\}$ is constant on $s=0$ (we assume $S$ is connected), and hence everywhere. It follows that $F|_S = A^{-1}$. Now, in the original geodesic Killing gauge we have that $\Omega$ satisfies $|\nabla \Omega|_g^2 =0$ and $\lie_{\eta}\Omega = - \Omega$. The function $v$ has the same properties and satisfies $\Omega = A^{-1} v$ on $\mc S$, so by Corollary \ref{cor_f1f2} one has $\Omega = A^{-1} v$ and $F= A^{-1}$ everywhere. Moreover, $G = \wh{G} F^2 = A^{-1}$ because $\wh{G} = A$. Performing a final constant rescaling of $v$ we finally arrive at
\begin{equation}
	\label{metric}
\Omega = v,\qquad g=2du dv +\mu.
\end{equation}
We emphasize that we have arrived at this expression starting with any $\Omega$ that belongs to the geodesic Killing gauge. The  RW coordinates $\{u,v\}$ in the final metric \eqref{metric} are fully determined in terms of $\Omega$. We call them \textit{adapted Rácz-Wald coordinates}. A conformal change within the geodesic Killing gauge has a highly non-trivial effect in the coordinates $\{u,v\}$ and also on $\mu$. However, at $\mc S$, the effect is simple. Any other $\wt{\Omega}$ in this gauge is uniquely parametrized by a positive function $\omega$ on $\mc S$ by means of $\wt{\Omega}|_{\mc S} = \omega \Omega|_{\mc S}$, and then $\tilde{\mu}|_{\mc S} = \omega^2 \mu|_{\mc S}$. So our construction keeps the full conformal freedom at $\mc S$. This is exactly the conformal freedom that exists in the original Fefferman-Graham construction.\\

Whenever $(g,\Omega)$ is quasi-Einstein one has that $\wt{g} = \Omega^{-2} g = v^{-2}(2du dv + \mu)$ is Ricci flat. Defining $\bar u \d  u$ and $\bar v \d v^{-1}$ one finds $$\wt g = -2d\bar u d\bar v + \bar v^2 \mu.$$ Therefore, this metric is exactly Ricci flat, admits a homothetic horizon ($\bar u=0$) and it is written in a double null coordinate system with $\Theta=1$, $q=0$ and $\slashed{g}= \bar v^2 \mu$. This proves that the class of ambient metrics of Theorem \ref{teo_rodnianski} is in one-to-one correspondence with the class of quasi-Einstein manifolds $(M,g,\Omega)$ that we have considered in this section. 

\begin{prop}
	\label{teorema0}
	There exists a one-to-one correspondence between straight, exact and regular ambient metrics and quasi-Einstein manifolds $(\mc M,g,\Omega)$ with $\lambda=0$ admitting an integrable conformal field $\lie_{\eta}g = 2\psi g$ satisfying $\lie_{\eta}\Omega = (\psi-1)\Omega$ and such that $\eta$ admits a bifurcate horizon where one of the horizons is $\scri$.
\end{prop}

This result relaxes the condition of $\eta$ being closed. As a consequence of the regularity in Theorem \ref{teo_rodnianski} and relation \eqref{partials}, the regularity of $g$ is as follows: For $\mf n=2$, $g$ is smooth. When $\mf n\ge 3$ is odd, $g$ is smooth everywhere except at $s=0$, and there exists smooth tensors $\{g^{(i)}_{\alpha\beta}\}_{i=0}^{\frac{\mf n-1}{2}}$ and $\wt{g}_{\alpha\beta}$ such that $$g_{\alpha\beta} = \sum_{i=0}^{\frac{\mf n-1}{2}} g_{\alpha\beta}^{(i)} s^i + s^{\frac{\mf n}{2}} \wt{g}_{\alpha\beta} + O\big(s^{\frac{\mf n+1}{2}}\big).$$ Finally, when $\mf n\ge 4$ is even, $g$ is smooth everywhere except at $s=0$, and there exists smooth tensors $\{g^{(i)}_{\alpha\beta}\}_{i=0}^{\frac{\mf n}{2}}$ and $\wt{g}_{\alpha\beta}$ such that $$g_{\alpha\beta} = \sum_{i=0}^{\frac{\mf n}{2}-1} g_{\alpha\beta}^{(i)} s^i + g_{\alpha\beta}^{(\mf n/2)}s^{\mf n/2} + \wt{g}_{\alpha\beta} s^{\mf n/2} \log(s) + O\big(s^{\frac{\mf n}{2}+1} \log(s)\big).$$ This allows us to identify the free data from the conformal picture as follows.

\begin{teo}
	\label{teorema}
Let $(\mc M,g,\Omega)$ be a regular quasi-Einstein manifold of dimension $\mf n+2$ that admits an integrable conformal Killing vector $\lie_{\eta} g = 2\psi g$ and a bifurcate horizon where one of the horizons is $\scri=\{\Omega=0\}\simeq \mc S\times\real$ and such that $\lie_{\eta}\Omega = (\psi-1)\Omega$. Write $(g,\Omega)$ in a geodesic Killing gauge and let $h$ be the metric of the bifurcation surface $\mc S$ and $\Psi$ the trace-free part of the term $s^{\mf n/2}$ in the expansion of $g$ in adapted Rácz-Wald coordinates. Then, $\Omega^{-2} g$ is the exact, straight and regular ambient metric with data $(\mc S,h,\Psi)$.
\end{teo}

It is also instructive to see how this free data appears by analysing the conformal Einstein equations order by order at $s=0$. Let us consider the $(A,B)$ and $(v,u)$ components of the quasi-Einstein equations of the metric \eqref{metricRWbeta=0}, that we write again here for completeness (see Appendix \ref{app})
\begin{align*}
\mc{Q}_{uv} & = \dfrac{v}{4\mf n(\mf n+1)}\Big( 2 R^{(h)}+ (\mf n-2) s  |\dot{\mu}|^2-2(\mf n-1)s\tr_{\mu}\ddot{\mu}+ \big(s\tr_{\mu}\dot{\mu} - 2(\mf n-1)\big)\tr_{\mu}\dot{\mu}\Big),\\
\mc{Q}_{AB} &= -\dfrac{v}{2\mf n} \Bigg(\big(\mf n-2-s\tr_{\mu}\dot{\mu}\big)\dot{\mu}_{AB} -2\big(R^{(h)}_{AB} -s(\dot{\mu}\cdot\dot{\mu})_{AB} +s\ddot{\mu}_{AB}\big)\\
&\qquad\qquad \, +\dfrac{1}{2(\mf n+1)} \Big(2 R^{(h)} -3s |\dot{\mu}|^2 +4s\tr_{\mu}\ddot{\mu} +(4+s\tr_{\mu}\dot{\mu} )\tr_{\mu}\dot{\mu}\Big)h_{AB}\Bigg).
\end{align*}
In order to study how the equations fix the geometry order by order at $s=0$ we start by solving $\mc{Q}_{uv}\st{s=0}{=} 0$ for $\tr_{\mu}\dot{\mu}$, which gives $\tr_{\mu}\dot{\mu} = \frac{1}{\mf n-1}R^{(h)}$. Inserting it into equation $\mc{Q}_{AB}\st{s=0}{=} 0$ we obtain $$(\mf n-2)\dot{\mu}_{AB} - 2R^{(h)}_{AB}-\dfrac{R^{(h)}}{\mf n-1}h_{AB} =0.$$ For $\mf n>2$ this equation fixes the tensor $\dot{\mu}_{AB}$ at $s=0$ to be twice the Schouten of $h_{AB}$, while for $\mf n=2$ this equation do not determine $\dot{\mu}_{AB}$ but holds automatically because in two dimensions $R^{(h)}_{AB} = \frac{1}{2} R^{(h)} h_{AB}$. Let us now analyse the equations order by order. To do that the strategy is to take $m$ derivatives of the equations w.r.t $s$ and keep track of the leading order terms. Dropping irrelevant global factors the result is (the symbol $\propto$ means proportionality with a non-zero factor)
\begin{align*}
\partial_s^{(m)}\big(v^{-1}\mc{Q}_{uv}\big) & \st{s=0}{\propto} \tr_{h}\mu^{(m+1)} + \mbox{l.o.t},\\
\partial_s^{(m)}\big(v^{-1}\mc{Q}_{AB}\big) &\st{s=0}{\propto} \big(\mf n-2-2m\big)\mu^{(m+1)}_{AB} + \dfrac{2m\tr_{h}\mu^{(m+1)}}{\mf n+1}h_{AB}+ \mbox{l.o.t},
\end{align*}
where ${\mu}^{(m)}\d \partial_s^{(m)}\mu$ and ``l.o.t'' stands for lower order terms. For every $1\le m<\frac{\mf n-2}{2}$ it is clear that the first equation determines the trace of $\mu^{(m+1)}_{AB}$, which inserted into the second gives the full $\mu^{(m+1)}_{AB}$. When $m=\frac{\mf n-2}{2}$ the trace-free part of $\mu^{(\mf n/2)}_{AB}$ cannot be determined from the equations, which makes the $\frac{\mf n}{2}$-term of the expansion free. Once such free data $\Psi_{AB}$ has been specified, one can continue determining the rest of the expansion.\\

That the free data $\Psi_{AB}$ must satisfy a divergence condition follows from Proposition \ref{teorema0}, as otherwise the ambient metric would not be straight. This can also be detected from the conformal viewpoint. Indeed, consider the $(u,A)$ components of the conformal equations, namely $$(\div_{\mu}{\mu})_A - \nabla_A\tr_{\mu}{\mu} = 0.$$ After taking $(\mf n/2-1)$ derivatives along $\partial_s$ one arrives to $$\div_h \Psi - \wt D \st{s=0}{=} 0,$$ where $\wt D$ is a tensor that only depends on $h_{AB}$. By Theorem \ref{teorema0}, this condition must be equivalent to \eqref{equationD} (i.e. $\wt{D}=D$).

\section{Conclusions and future work}

In this paper, we have shown that any straight ambient metric admits a conformal completion with a bifurcate Killing horizon with integrable Killing one-form, where one of the branches corresponds to $\scri$ and the other to the original homothetic horizon. This implies that the transverse expansion of the metric at null infinity can be related to the one at the homothetic horizon. In particular, the free data in the Fefferman–Graham construction appears at the same order as the gravitational radiation at $\scri$. We then established a one-to-one correspondence between exact, straight ambient metrics and solutions of the conformal Einstein equations that admit a bifurcate conformal Killing horizon with integrable Killing one-form and whose conformal factor satisfies a suitable condition. Notably, this result identifies the Fefferman–Graham ambient metric as the unique conformal spacetime with these properties. We believe this result is relevant for at least two reasons: first, because it characterizes the ambient metric from a conformal geometric perspective; and second, because it relaxes the requirement for the homothety to be exact as a one-form.\\

The results of this paper open the door for our next work \cite{Mio7}, where we will employ the general identities developed in \cite{Mio3,Mio4} to provide a fully geometric characterization of asymptotically flat spacetimes. As we will see, the gravitational degrees of freedom for a general asymptotically flat spacetime of arbitrary dimension appear at the same order as the free data in the ambient metric. This suggests a strong connection between the ambient metric and a geometric definition of radiation that we plan to analyse further in future works. Moreover, in \cite{Mio7} we will show that one intrinsic obstruction for the smoothness of $\scri$ is strongly related to the presence of a non-vanishing obstruction tensor $\mc{O}_{AB}$ at the cuts of $\scri$.\\


Another promising direction for future work is to analyze the conformal freedom in the Fefferman-Graham construction also from the conformal perspective, that is, to determine which data $(\omega^2h,\Psi')$ give rise to the same ambient metric as $(h,\Psi)$. This analysis could shed light on the general (and complicated) question of when two asymptotically flat spacetimes in arbitrary dimension with conformally related universal structures are equivalent, or in other words, how radiation behaves under conformal transformations.

\section*{Acknowledgements}
This work has been supported by Projects PID2024-158938NB-I00 (Spanish Ministerio de Ciencia e Innovación and FEDER ``A way of making Europe''). M. Mars acknowledges financial support under projects SA097P24 (JCyL) and RED2022-134301-T funded by MCIN/AEI/10.13039/ 501100011033. G. Sánchez-Pérez also acknowledges support of the PhD. grant FPU20/03751 from Spanish Ministerio de Universidades.

\begin{appendices}
	
	\section{Quasi-Einstein equations in Rácz-Wald coordinates}
	\label{app}
In this appendix we write down explicitly the quasi-Einstein equations associated to the metric \eqref{metricRWbeta=0} in Rácz-Wald coordinates. More specifically we compute the components of the tensor $\mc{Q}\d \mf  \hess v +v \sch_g$. The computation has been done using the \texttt{xAct} package of \texttt{Mathematica}. We introduce the variable $s\d uv$ and we denote a derivative w.r.t. $s$ with a dot. 
\begin{align*}
\mc{Q}_{vv} & = \dfrac{u}{4\mf{n}} \left(2\big(s \tr_{\mu}\dot{\mu}-2\mf n   \big)\dfrac{\dot{G}}{G}+ s\big(|\dot{\mu}|^2- 2\tr_{\mu}\ddot{\mu}\big)\right),\\
\mc{Q}_{uv} & = \dfrac{v}{4\mf{n}(\mf n+1)}\Big(-4\mf n \dfrac{\dot{G}}{G} -2G R^{(\mu)} +4\mf n s \dfrac{\dot{G}^2}{G^2}-4\mf ns\dfrac{\ddot{G}}{G}+ (\mf n-2) s  |\dot{\mu}|^2-2(\mf n-1)s\tr_{\mu}\ddot{\mu}\\
&\qquad \qquad \qquad + \big(s\tr_{\mu}\dot{\mu} - 2(\mf n-1)\big)\tr_{\mu}\dot{\mu} - 2(\mf n-1) \square_{\mu}G - G^{-1} |\nabla G|_{\mu}^2\Big),\\
\mc{Q}_{uu} & = \dfrac{v^3}{4\mf n}\left(|\dot{\mu}|^{2} - 2\tr_{\mu}\ddot{\mu}+2\dfrac{\dot{G}}{G}\tr_{\mu}\dot{\mu}\right),\\
\mc{Q}_{vA} & = \dfrac{1}{4\mf{n}}\left(\left(-2\mf n  + 2s\dfrac{\dot G}{G} +s\tr_{\mu}\dot{\mu}\right)\dfrac{\nabla_A G}{G} - 2s\left(\dfrac{\nabla_A\dot G}{G} +  \big(\nabla_A\tr_{\mu}\dot{\mu} - (\div_{\mu}\dot{\mu})_A\big)\right)\right),\\
\mc{Q}_{uA} & = \dfrac{v^2}{4\mf{n}} \left(\left(2\dfrac{\dot G}{G}+\tr_{\mu}\dot{\mu}\right)\dfrac{\nabla_A G}{G} -2 \left(\dfrac{\nabla_A\dot{G}}{G}+\big(\nabla_A\tr_{\mu}\dot{\mu}-(\div_{\mu}\dot{\mu})_A\big)\right)\right),\\
\mc{Q}_{AB} & = \dfrac{v}{4\mf{n}G} \Bigg(\Big(2\big(\mf n-2-s\tr_{\mu}\dot{\mu}\big)\dot{\mu}_{AB} +2\big(2G R^{(\mu)}_{AB} +2s\big(\dot{\mu}\cdot \dot{\mu}\big)_{AB} -2s\ddot{\mu}_{AB}  -2\nabla_A\nabla_B G + G^{-1}\nabla_AG\nabla_BG\big) \Big)\\
&  \qquad\qquad +\dfrac{1}{\mf n+1} \Bigg(4\dfrac{\dot G}{G} -2G R^{(\mu)}  +4s\left(\dfrac{\ddot{G}}{G}-\dfrac{\dot{G}^2}{G^2}\right) -3s |\dot{\mu}|^2 +4s\tr_{\mu}\ddot{\mu} +(4+s\tr_{\mu}\dot{\mu} )\tr_{\mu}\dot{\mu} \\
&\quad\qquad\qquad\qquad +4\square_{\mu} G - G^{-1} |\nabla G|^2\Bigg)\mu_{AB}\Bigg),
\end{align*}
where $\tr_{\mu}$ is the trace w.r.t. $\mu$ and $(\dot\mu\cdot\dot\mu)_{AB}\d \mu^{CD}\dot\mu_{AC}\dot\mu_{BD}$. 
Particularizing the equations for $G=-1$,
\begin{align*}
	\mc{Q}_{vv} & = \dfrac{us}{4\mf{n}} \big(|\dot{\mu}|^2 - 2\tr_{\mu}\ddot{\mu}\big),\\
	\mc{Q}_{uv} & = \dfrac{v}{4\mf{n}(\mf n+1)}\Big( 2 R^{(\mu)}+ (\mf n-2) s  |\dot{\mu}|^2-2(\mf n-1)s\tr_{\mu}\ddot{\mu}+ \big(s\tr_{\mu}\dot{\mu} - 2(\mf n-1)\big)\tr_{\mu}\dot{\mu} \Big),\\
	\mc{Q}_{uu} & = \dfrac{v^3}{4\mf{n}}\left(|\dot{\mu}|^2 - 2\tr_{\mu}\ddot{\mu}\right),
\end{align*}
\begin{align*}
	\mc{Q}_{vA} & = -\dfrac{s}{2\mf{n}}\left(\nabla_A\tr_{\mu}\dot{\mu} - (\div_{\mu}\dot{\mu})_A\right),\\
	\mc{Q}_{uA} & = -\dfrac{v^2}{2\mf{n}} \left( \nabla_A\tr_{\mu}\dot{\mu} - (\div_{\mu}\dot{\mu})_A\right),\\
	\mc{Q}_{AB} & = -\dfrac{v}{2\mf{n}} \Bigg(\big(\mf n-2-s\tr_{\mu}\dot{\mu}\big)\dot{\mu}_{AB} -2\big( R^{(\mu)}_{AB} -s(\dot{\mu}\cdot\dot{\mu})_{AB} +s\ddot{\mu}_{AB}\big) \\
	&\qquad\qquad  +\dfrac{1}{2(\mf n+1)} \Big(2 R^{(\mu)} -3s |\dot{\mu}|^2 +4s\tr_{\mu}\ddot{\mu} +(4+s\tr_{\mu}\dot{\mu} )\tr_{\mu}\dot{\mu}\Big)\mu_{AB}\Bigg).
\end{align*}

\end{appendices}

\begingroup
\let\itshape\upshape

\renewcommand{\bibname}{References}
\bibliographystyle{acm}
\bibliography{biblio} 

\end{document}